\documentclass[aps,prd,onecolumn,groupedaddress,showpacs,nofootinbib,amssymb]{revtex4}
\usepackage[dvips]{graphicx}
\usepackage{amssymb}
\usepackage{amsmath}
\usepackage{graphicx}
\usepackage{amsfonts}
\usepackage{bm}

\begin{document}

\title{Loop Quantum Cosmology Corrected Gauss-Bonnet Singular Cosmology}
\author{
K. Kleidis,$^{1}$\,\thanks{kleidis@teiser.gr}
V.K. Oikonomou$^{1}$,\,\thanks{v.k.oikonomou1979@gmail.com}}
\affiliation{
$^{1)}$ Department of Mechanical Engineering\\ Technological
Education Institute of Central Macedonia \\
62124 Serres, Greece
}

\begin{abstract}
In this work we investigate which Loop Quantum Cosmology corrected Gauss-Bonnet $F(\mathcal{G})$ gravity can realize two singular cosmological scenarios, the intermediate inflation and the singular bounce scenarios. The intermediate inflation scenario has a Type III sudden singularity at $t=0$, while the singular bounce has a soft Type IV singularity. By using perturbative techniques, we find the holonomy corrected $F(\mathcal{G})$ gravities that generate at leading order the aforementioned cosmologies and we also argue that the effect of the holonomy corrections is minor to the power spectrum of the primordial curvature perturbations of the classical theory.
\end{abstract}

\pacs{04.50.Kd, 95.36.+x, 98.80.-k, 98.80.Cq,11.25.-w}

\maketitle



\def\pp{{\, \mid \hskip -1.5mm =}}
\def\cL{\mathcal{L}}
\def\be{\begin{equation}}
\def\ee{\end{equation}}
\def\bea{\begin{eqnarray}}
\def\eea{\end{eqnarray}}
\def\tr{\mathrm{tr}\, }
\def\nn{\nonumber \\}
\def\e{\mathrm{e}}

\section*{Introduction}

The primordial era of our Universe is one of the most mysterious challenges of modern cosmology. Up to date we can speculate only for the era during and after the inflationary epoch \cite{inflation1,inflation2,inflation3,inflation4,reviews1,reviews2,reviews3,reviews4}, however, it is still vague whether the inflationary era occurred in the first place, since various alternative scenarios to the standard inflationary scenario exist, such as the bounce cosmology scenarios \cite{Brandenberger:2012zb,Brandenberger:2016vhg,Battefeld:2014uga,Novello:2008ra,Cai:2014bea,deHaro:2015wda,Lehners:2011kr,Lehners:2008vx,Nojiri:2016ygo,Odintsov:2015ynk,Odintsov:2014gea}. Also there are scenarios which combine the inflationary and bounce cosmology era, see for example \cite{Liu:2013kea,Piao:2003zm}. In all the aforementioned cases, the era before the inflationary epoch is quite mysterious and it is believed that the quantum theory of gravity governs this era, except for the scenarios that the bounce and inflationary cosmologies are combined \cite{Liu:2013kea,Piao:2003zm}. One appealing quantum theory of gravity that may offer interesting information to the primordial era is Loop Quantum Cosmology (LQC) \cite{LQC1,LQC2,LQC3,LQC4,LQC5,LQC6,LQC19,singh,Haro:2017mir,Salo:2016dsr,Amoros:2014tha,Amoros:2013nxa,Bamba:2012ka,Mielczarek:2009kh,Mielczarek:2008zz,Mielczarek:2007zy,Oikonomou:2017mlk,Oikonomou:2016phk,Odintsov:2015uca}, in the context of which many inflationary and bounce cosmology scenarios may be realized \cite{Amoros:2014tha,Amoros:2013nxa,Bamba:2012ka,Mielczarek:2009kh,Mielczarek:2008zz,Mielczarek:2007zy,Oikonomou:2017mlk,Oikonomou:2016phk,Odintsov:2015uca}. In some cases it is possible to find the imprint of the LQC on various physical quantities during the inflationary era. Particularly, this can be done by using a perturbative expansion of the effects of LQC on classical theories of gravity, and through these, the impact of the primordial era on present day physics can be revealed. In this line of research, we shall use the framework of Ref. \cite{Haro:2015oqa}, and we shall investigate which LQC-corrected Gauss-Bonnet $F(\mathcal{G})$ \cite{Haro:2015oqa} gravity may describe a well-known singular inflation scenario, the intermediate inflation scenario \cite{Barrow:1990td,Barrow:1993zq,Barrow:2006dh,Barrow:2014fsa,Oikonomou:2017isf,Oikonomou:2017brl}, and also we apply the formalism for generating the singular bounce. Modified Gauss-Bonnet gravity \cite{Li:2007jm,Nojiri:2005jg,Nojiri:2005am,Cognola:2006eg,Elizalde:2010jx,Izumi:2014loa,Oikonomou:2016rrv,Oikonomou:2015qha,Escofet:2015gpa,Makarenko:2017vuk,new2,Makarenko:2016jsy} is known to provide a successful theoretical framework in the context of which various cosmological scenarios may be described, hence in this work we study the holonomy corrected version of the theory. In addition, the intermediate inflation scenario is a singular type of inflation \cite{Barrow:2015ora,Nojiri:2015fra,Odintsov:2015gba,Oikonomou:2015qfh}, and particularly it is a Type III finite time singularity, following the classification of \cite{Nojiri:2005sx}. This was also pointed out in Ref. \cite{Oikonomou:2017brl}, and a similar analysis was performed in the context of $F(T)$ gravity. Also the singular bounce is a particular bouncing cosmology example, in which a Type IV singularity occurs at the origin $t=0$ \cite{Nojiri:2015fra,Odintsov:2015gba,Oikonomou:2015qfh}. In the singular bounce case, the Universe may smoothly pass through the singularity without having any crushing type catastrophic event. The dynamics of the bounce and inflationary power spectrum is affected though, as was shown in Refs. \cite{Odintsov:2015gba,Oikonomou:2015qfh}. In our study, the metric shall be assumed to be a flat Friedman-Robertson-Walker (FRW) metric, and we shall find which holonomy corrected LQC $F(\mathcal{G})$ gravity may generate the intermediate inflationary scenario and the singular bounce. In some sense, we are seeking for the quantum effects of LQC on the classical evolution of $F(\mathcal{G})$ gravity, and to our opinion, the perturbative study we perform is the first step towards seeing the quantum effects on a classical level, in the context of $F(\mathcal{G})$ gravity. However, the effects on the observational indices of the LQC corrections are minor, since these are subleading order effects, which have a perturbation expansion origin.

This paper is organized as follows: In section I we briefly review the LQC holonomy corrected $F(\mathcal{G})$ gravity formalism, and by using this formalism, in section III we investigate which LQC holonomy corrected $F(\mathcal{G})$ gravity may generate the Type III singular inflation scenario and the Type IV singular bounce scenario, by focusing on the first correction term at leading order in the perturbative expansion. Finally, the conclusions follow in the end of the paper.

Before we start our presentation, we briefly discuss the geometric framework of our analysis. Particularly, we shall assume that the background geometry is described by a flat FRW metric, with line element,
\be
\label{metricfrw} ds^2 = - dt^2 + a(t)^2 \sum_{i=1,2,3}
\left(dx^i\right)^2\, ,
\ee
with $a(t)$ being the scale factor of our Universe. In addition, we shall assume that the spacetime connection is a torsion-less, metric compatible affine connection, the Levi-Civita connection. Finally, the Gauss-Bonnet invariant for the FRW metric has the following form,
\begin{equation}\label{gaussbonnetinvariant}
\mathcal G=24 H(t)^2 \left(H'(t)+H(t)^2\right)\, ,
\end{equation}
with $H(t)$ denoting as usual the Hubble rate of the Universe $H(t)=\dot{a}/a$.

\section{Overview of LQC-corrected Gauss-Bonnet Gravity}

The Gauss-Bonnet theory of gravity is a Jordan frame theory and no equivalence to standard general relativity exists. The LQC corrections can be introduced in this theory by using the formalism of Ref. \cite{Haro:2015oqa}, according to which, holonomy corrections can be introduced in the $F({\mathcal G})$ theory by making the following replacement
$\beta\rightarrow \frac{\sin(\lambda \beta)}{\lambda}$ or equivalently $p_V\rightarrow -\frac{2\sin(\lambda \beta)}{\lambda\gamma}$, in the Hamiltonian of the theory, which is,
\begin{eqnarray}\label{A7}
 {\mathcal H}_{grav}(V,p_V)=-\frac{3}{4}p_V^2V=-3H^2V\, ,
 \end{eqnarray}
where $\beta=-\frac{\gamma}{2}p_V=\gamma H$, $\gamma$ is the Barbero-Immirzi parameter, and $V$ is the world volume $V=a^3$. The Hamiltonian (\ref{A7}), in terms of the parameter $\beta$
can be written ${\mathcal H}_{grav}(V,\beta)=-\frac{3\beta^2}{\gamma^2}V$, and the vacuum non-holonomy corrected $F(\mathcal{G})$ gravity Raychauduri equation is written as follows,
\begin{equation}\label{raychauduri}
\dot{p_{ V}}=-\frac{\partial{\mathcal H}_{grav}}{\partial V}\, .
\end{equation}
The holonomy corrections can be obtained if the momentum $-2\frac{\beta}{\gamma}$ is replaced by $-\frac{2\sin(\lambda \beta)}{\lambda\gamma}$, which in the case that the variables $(V,{\mathcal G},p_V,p_{\mathcal G})$ are used, the replacement reads, $p_V\rightarrow -\frac{2\sin(\lambda \beta)}{\lambda\gamma}$, in the classical Hamiltonian, which is,
\begin{eqnarray}\label{A5}
 {\mathcal H}_{grav}(V,{\mathcal G},p_V,p_{\mathcal G})\equiv \dot{V}p_V+\dot{{\mathcal G}}p_{\mathcal G}-\tilde{\mathcal L}(V,\dot{V},{\mathcal G},\dot{{\mathcal G}})\nonumber\\
 =\frac{3}{V}\left(\frac{V^2p_{\mathcal G}}{4F''({\mathcal G})} \right)^{2/3}-3p_V\left(\frac{V^2p_{\mathcal G}}{4F''({\mathcal G})} \right)^{1/3}+\frac{V}{2}\left({\mathcal G}F'({\mathcal G})-F({\mathcal G})\right).
\end{eqnarray}
In principle, there are multiple alternative ways to introduce holonomy corrections, depending on the choices of variables that are used for the formulation of the $F({\mathcal G})$ theory. Then, in the general case that matter fluids are also taken into account, in which case the term ${\mathcal H}_{matter}=\rho V$ must be added in the Hamiltonian, the replacement $p_V\rightarrow -\frac{2\sin(\lambda \beta)}{\lambda\gamma}$ in the classical Hamiltonian leads to the following Hamiltonian,
\begin{eqnarray}\label{A12}
 {\mathcal H}_{hol}(V,{\mathcal G},p_V,p_{\mathcal G})
 =\frac{3}{V}\left(\frac{V^2p_{\mathcal G}}{4F''({\mathcal G})} \right)^{2/3}
 \nonumber\\+\frac{6\sin(\lambda \beta)}{\lambda\gamma}\left(\frac{V^2p_{\mathcal G}}{4F''({\mathcal G})} \right)^{1/3}+\frac{V}{2}\left({\mathcal G}F'({\mathcal G})-F({\mathcal G})\right)
 +\rho V\, ,
\end{eqnarray}
and in effect, the Hamilton equations are cast as follows,
\begin{eqnarray}\label{A13}
 \dot{V}=-\frac{\gamma}{2}\frac{\partial {\mathcal H}_{hol} }{\partial\beta};\quad \dot{\mathcal G}=\frac{\partial {\mathcal H}_{hol} }{\partial p_{\mathcal G}}\, .
\end{eqnarray}
By taking into account the Hamiltonian constraint ${\mathcal H}_{hol}(V,{\mathcal G},p_V,p_{\mathcal G})=0$, we obtain,
\begin{eqnarray}\label{A14}
 H=-\cos(\lambda\beta)\tilde{p}_{\mathcal G}^{\frac{1}{3}}\nonumber\\
 \dot{\mathcal G}=\frac{1}{2F''({\mathcal G})}\left(\tilde{p}_{\mathcal G}^{-\frac{1}{3}}+\tilde{p}_{\mathcal G}^{-\frac{2}{3}} \frac{\sin(\lambda \beta)}{\lambda\gamma} \right)
 \nonumber\\
 3\tilde{p}_{\mathcal G}^{\frac{2}{3}}+\tilde{p}_{\mathcal G}^{\frac{1}{3}} \frac{6\sin(\lambda \beta)}{\lambda\gamma}+\frac{1}{2}\left({\mathcal G}F'({\mathcal G})-F({\mathcal G})\right)
 +\rho=0,
\end{eqnarray}
where $\tilde{p}_{\mathcal G}=\frac{p_{\mathcal G}}{4VF''({\mathcal G})}$. The dynamical evolution of the cosmological system is determined by Eq. (\ref{A14}) in conjunction with $G(H, {\mathcal G},\dot{\mathcal G},\rho)=0$, with the conservation equation $\dot{\rho}=-3H(\rho+P)$ and with the equation of state $P=P(\rho)$ for the matter fields. By combining the three equations of Eq. (\ref{A14}), we obtain the LQC modified Friedmann equation
in the context of $F(\mathcal G)$ gravity, which is,
\begin{eqnarray}\label{A19}
4\tilde{p}_{\mathcal G}^{\frac{4}{3}}(F''({\mathcal G}))^2 \dot{\mathcal G}^2
=\frac{\rho_c}{3}\left(1-\frac{H^2}{\tilde{p}_{\mathcal G}^{\frac{2}{3}}} \right)-\frac{1}{6}({\mathcal G}F'({\mathcal G})-F({\mathcal G}))-
\frac{\rho}{3}.
\end{eqnarray}
Combining the above, the dynamical evolution in LQC holonomy corrected $F(\mathcal G)$ Gauss-Bonnet gravity, is determined by the following equations,
\begin{eqnarray}\label{A20}
 4\tilde{p}_{\mathcal G}^{\frac{4}{3}}(F''({\mathcal G}))^2 \dot{\mathcal G}^2
=\frac{\rho_c}{3}\left(1-\frac{H^2}{\tilde{p}_{\mathcal G}^{\frac{2}{3}}} \right)-\frac{1}{6}({\mathcal G}F'({\mathcal G})-F({\mathcal G}))-
\frac{\rho}{3}\nonumber\\
\dot{\rho}=-3H(\rho+P),
\end{eqnarray}
where the parameter $\rho_c$ is the critical density, which measures the quantum nature of the cosmological dynamics. To be more specific, the critical density $\rho_c$ controls the effects of LQC in the classical evolution, and if the limit $\rho_c\to \infty$ is taken, the classical gravity limit is approached, and the first quantum corrections may appear near the classical limit. In effect, if $\rho$ is finite and large but bounded from above, the manifestation of the quantum effects can be seen at a perturbative level in terms of this critical density. This is exactly the research line we shall adopt in this article. In order to finding the LQC corrections to the standard $F(\mathcal{G})$ gravity, we rewrite Eq. (\ref{A20}) as follows,
\begin{equation}
24(F^{\prime\prime}(\mathcal{G})\Dot{\mathcal{G}})^2{\bar p}^2_{\mathcal{G}}(\mathcal{G})
-2\rho_c\left(1-\frac{H^2}{{\bar p}_{\mathcal{G}}(\mathcal{G})}\right)
+A(\mathcal{G})=0,
\label{fg1}
\end{equation}
where $\bar{p}_{\mathcal{G}}(\mathcal{G})$ is,
$$\bar{p}_{\mathcal{G}}(\mathcal{G})\equiv
\frac{1}{6}\left(4\rho_c-A(\mathcal{G})
-2\sqrt{2\rho_c}\sqrt{2\rho_c-A(\mathcal{G})-6H^2}\right).$$
In the limit $\rho_c\to\infty$, the quantity ${\bar p}_{\mathcal{G}}(\mathcal{G})$
can be written as a perturbative sum,
$${\bar p}_{\mathcal{G}}(\mathcal{G})=
H^2+\frac{1}{48}\left(A(\mathcal{G})+6H^2\right)^2\frac{1}{\rho_c}
+\frac{1}{192}\left(A(\mathcal{G})+6H^2\right)^3\frac{1}{\rho_c^2}
+o\left(\frac{1}{\rho_c^3}\right).$$
and in effect, Eq. (\ref{fg1}) can be cast in the following way,
\begin{multline}
576H^6(F^{\prime\prime}(\mathcal{G})\Dot{\mathcal{G}})^2
-\left(A(\mathcal{G})-6H^2\right)^2
+\\+
\varepsilon\frac{\left(A(\mathcal{G})+6H^2\right)^2}{48H^2}
\left(A^2(\mathcal{G})-36H^4+1152H^6(F^{\prime\prime}(\mathcal{G})\Dot{\mathcal{G}})^2\right)+\\
+\varepsilon^2\frac{\left(A(\mathcal{G})+6H^2\right)^3}{2304H^4}\times \\  \times
\left(A^3(\mathcal{G})-6A^2(\mathcal{G})H^2+432H^6-576H^6(A(\mathcal{G}+30H^2))(F^{\prime\prime}(\mathcal{G})\Dot{\mathcal{G}})^2\right)+\ldots=0.
\label{fg2}
\end{multline}
As we already mentioned, if the limit $\rho_c\rightarrow\infty$ is taken, the equation above becomes the classical Friedman equation for $F(\mathcal{G})$ gravity. In effect, the parameter $\rho_c$ can be used as a perturbative parameter. We define $\varepsilon=1/\rho_c$, so in the limit $\rho_c\rightarrow\infty$, the parameter $\varepsilon$ takes small values, and thus this is the perturbative parameter we seek, which depends on the critical density $\rho_c$. In order to find the first quantum corrections originating to the LQC corrected theory, we expand the $F(\mathcal{G})$ gravity as follows,
\begin{equation}
F(\mathcal{G})=\sum\limits_{k=0}^{\infty}\varepsilon^kF_k(\mathcal{G}).
\label{fg3}
\end{equation}
In effect, the tree level ($\varepsilon=0$) $F(\mathcal{G})$ gravity, that realizes a specifically chosen cosmological evolution with Hubble rate $H$, is determined by solving the following differential equation,
\begin{equation}
24H^3F_0^{\prime\prime}(\mathcal{G})\Dot{\mathcal{G}}
-\mathcal{G}F_0^{\prime}(\mathcal{G})
+F_0(\mathcal{G})
+6H^2
-2\rho(t)=0\, .
\label{fg4}
\end{equation}
where we denoted as $F_0(\mathcal{G})$ the zeroth order contribution. Accordingly, at first order in the $\varepsilon$-perturbative expansion, the $F_1(\mathcal{G})$ gravity reads,
\begin{equation}
24H^3F_1^{\prime\prime}(\mathcal{G})\Dot{\mathcal{G}}
-\mathcal{G}F_1^{\prime}(\mathcal{G})
+F_1(\mathcal{G})=
-18H^4(1+2HF_0^{\prime\prime}(\mathcal{G})\Dot{\mathcal{G}})^2(1+6HF_0^{\prime\prime}
(\mathcal{G})\Dot{\mathcal{G}})\, .
\label{fg5}
\end{equation}
In the following we shall only be interested in finding the zeroth and first order contributions $F_0(\mathcal{G})$ and $F_1(\mathcal{G})$ respectively, in the $\varepsilon$-perturbative expansion of Eq. (\ref{fg3}), given the Hubble rate of the cosmological evolution. Essentially, the differential equations (\ref{fg4}) and (\ref{fg5}), given the Hubble rate, constitute a reconstruction method, which we shall use in the following, in order to realize a singular evolution. Since we shall be interested in the vacuum Gauss-Bonnet gravity case, we shall assume that the matter fluids do not affect the dynamical evolution of the cosmological system, and therefore $\rho(t)=0$.

\section{Intermediate Inflation and Singular Bounce from LQC-corrected Gauss-Bonnet Gravity}

Having described in the previous section the LQC $f(\mathcal{G})$ gravity formalism, in this section we shall be interested in realizing a singular inflationary cosmology and a singular bouncing cosmology, in which a finite time singularity of Type III or Type IV occurs at the origin $t=0$. Particularly, we shall be interested in the intermediate inflation scenario \cite{Barrow:1990td,Barrow:1993zq,Barrow:2006dh,Barrow:2014fsa}, which was firstly introduced by Barrow in \cite{Barrow:1990td}, which as was shown in Ref. \cite{Oikonomou:2017brl} corresponds to the expanding phase of a Type III singular bounce cosmological evolution. The scale factor and the Hubble rate of the intermediate inflation scenario have the following form \cite{Barrow:1990td},
\begin{equation}\label{bambabounce}
a(t)=e^{A\,t^n},\,\,\, H(t)=A n t^{n-1}\, ,
\end{equation}
with $0<n<1$ and also with $A>0$ and of course $t>0$. Accordingly, the singular bounce is described by the same functional dependence as in Eq. (\ref{bambabounce}), but in this case, $n>2$ and $-\infty<t<\infty$. Also in the latter case, $n$ must be chosen in such a way so that complex values in the Hubble rate are avoided, for example $n=\frac{2n}{2m+1}>2$ and being an integer with $m>0$. In order to understand the finite-time singularity structure of the intermediate inflation scenario and of the singular bounce, let us recall in brief the classification of finite-time singularities following Refs.~\cite{Nojiri:2005sx}:
\begin{itemize}
\item Type I (``Known as Big Rip''): This is a crushing type singularity, in which all the physical quantities defined on the constant-time three dimensional spacelike hypersurface diverge, that is, $a \to \infty$, $\rho_\mathrm{eff} \to \infty$, and
$\left|p_\mathrm{eff}\right| \to \infty$.
\item Type II (``Known as Sudden Singularity, or pressure singularity''): In this case,  only the effective pressure diverges, that is, $\left|p_\mathrm{eff}\right| \to \infty$. This type of singularity has been extensively studied in Refs. \cite{Nojiri:2004ip,Nojiri:2005sr,Barrow:2004he,Barrow:2004hk,Barrow:2004xh}.
\item Type III : In this singularity type, only the scale factor remains finite, but the rest of the physical quantities diverge, that is, $\rho_\mathrm{eff} \to \infty$,
$\left|p_\mathrm{eff}\right| \to \infty$. This type of singularity was discovered in Ref. \cite{Nojiri:2004pf}.
\item Type IV : This is a non-crushing type singularity, in which the Universe may smoothly pass through it, without any catastrophic effects. In this case, all the physical quantities remain finite at the spacelike hypersurface $t=t_s$ defined by the time instance $t_s$ that the singularity occurs. In this case, divergences may occur in the higher derivatives of the Hubble rate. The effects of this type of singularities on the dynamical evolution of inflationary theories, were extensively studied in Refs. \cite{Nojiri:2015fra,Odintsov:2015gba,Odintsov:2015zza}, see also \cite{Oikonomou:2015qfh,Brevik:2016kuy} for some alternative implications.
\end{itemize}
From the above classification, it is easy to see that the singularity type of the intermediate inflation scenario (\ref{bambabounce}) is Type III, since the first derivative of the Hubble rate diverges for $0<n<1$, and in addition, the singular bounce has a Type IV singularity at the origin $t=0$.

Let us now proceed to the investigation in order to find which perturbative LQC-corrected $F(\mathcal{G})$ gravity may generate the evolution (\ref{bambabounce}) at leading order, in both the cases $0<n<1$ (intermediate inflation) or $n>2$ (singular bounce). We start off with the intermediate inflation scenario, so let us first find the classical contribution to the $F(\mathcal{G})$ gravity, so let us solve the differential equation (\ref{fg4}) for the cosmic evolution chosen as in Eq. (\ref{bambabounce}). The full differential equation is not possible to be solved analytically, so we need to find an approximate expression at leading order. Since we are interested in early times, the cosmic time is small, and in effect, the Gauss-Bonnet invariant is approximately equal to,
\begin{equation}\label{approximategaussbonnet}
\mathcal{G}=24 A^3 n^3 (n-1) t^{3 n-4}\, ,
\end{equation}
which means that since $0<n<1$, as $t\to 0$, $\mathcal{G}\to \infty$. Therefore, the leading order contributions in $\mathcal{G}$ should be chosen by taking into account the largest orders in $\mathcal{G}$. By doing so, the differential equation (\ref{fg4}) reads,
\begin{equation}\label{diffeqnnewleadingordering}
\frac{B}{\mathcal{G}^{\gamma }}+\mathcal{G}^2 K F_0''(\mathcal{G})-\mathcal{G} F_0'(\mathcal{G})+F_0(\mathcal{G})=0\, ,
\end{equation}
where the parameters $\gamma$, $B$ and $K$ are equal to,
\begin{align}\label{parametersBandK}
& \gamma=\frac{2 (n-1)}{4-3 n},\\ \notag &
B=6 A^2 n^2 \left(24 A^3 n^4-24 A^3 n^3\right)^{\frac{2 (n-1)}{4-3 n}},\\ \notag &
K=\frac{(3 n-4) ((4 A+3) n-4)}{n-1}\, .
\end{align}
By solving the differential equation (\ref{diffeqnnewleadingordering}), we obtain the classical contribution to the $F(\mathcal{G})$ gravity, which is,
\begin{equation}\label{classicalcontribution}
F_0(\mathcal{G})=-\frac{B \mathcal{G}^{-\gamma }}{(\gamma +1) (\gamma  K+1)}+c_1 \mathcal{G}^{1/K}+c_2 \mathcal{G}\, ,
\end{equation}
and it complies with the leading order result of Ref. \cite{Oikonomou:2017brl} for the classical $F(\mathcal{G})$ gravity. Also, for the classical case, when $0<n<1$, if one considers an $F(\mathcal{G})$ gravity in vacuum, with action,
\begin{equation}\label{actionfggeneral}
\mathcal{S}=\frac{1}{2\kappa^2}\int \mathrm{d}^4x\sqrt{-g}\left (
R+F(\mathcal{G})\right )\, ,
\end{equation}
where $\kappa^2=1/M_{pl}^2$, and also $M_{pl}=1.22\times 10^{19}$GeV, the power spectrum of the primordial curvature perturbations can be easily calculated, by following the technique used in Ref. \cite{Oikonomou:2015qha}, and the result is,
\begin{equation}\label{powerspectrumfinal}
\mathcal{P}_R\sim k^{\frac{7}{2}+3+\frac{\left(2-2 (n-1) +(n-1)
^2\right) \mu }{2 (n )}}\, ,
\end{equation}
where $\mu=-11/n$. Having the power spectrum at hand, enables us to calculate the spectral index of the primordial curvature perturbations, which is in this case,
\begin{equation}\label{rebel}
 n_s-1\equiv\frac{d\ln\mathcal{P}_{\mathcal{R}}}{d\ln
k}\simeq 1-\frac{11}{2n^2}\, .
\end{equation}
The 2015 Planck observational data \cite{Ade:2015lrj} constrain the spectral index $n_s$ as follows,
\begin{equation}
\label{planckdata} n_s=0.9644\pm 0.0049\, ,
\end{equation}
so the spectral index in Eq. (\ref{rebel}) can be compatible with the observations only when  $12.9<n<13.7$. Hence, the classical $F_0(\mathcal{G})$ intermediate inflation is not compatible with the observational data. Due to the fact that the LQC contribution is subleading, we conclude that the LQC corrected theory cannot render the theory compatible with data, unless the perturbation parameter $\rho_c$ takes small values, which would break the perturbation expansion. At this point, let us calculate the first LQC quantum correction to the $F(\mathcal{G})$ theory, which can be found if we solve the differential equation (\ref{fg5}), by also using the solution (\ref{classicalcontribution}). The differential equation (\ref{fg5}) at leading order reads,
\begin{equation}\label{firstperturbativeorder}
-\Gamma \mathcal{G}^{\delta}+\mathcal{G}^2 K F_1''(\mathcal{G})-\mathcal{G} F_1'(\mathcal{G})+F_1(\mathcal{G})=0\, ,
\end{equation}
where the parameters $\Gamma$ and $\delta$ are,
\begin{align}\label{gammaparameterxekarfotina}
& \Gamma=\frac{23887872 A^{16} B^3 \gamma ^3 (n-1)^3 n^{16} (3 n-4)^2 \left(24 A^3 n^4-24 A^3 n^3\right)^{\frac{4 (n-1)}{4-3 n}+\frac{8 n}{4-3 n}-\frac{23}{4-3 n}}}{(\gamma  K+1)^3},
\\ \notag & \delta=-3 \gamma -\frac{4 (n-1)}{4-3 n}-\frac{12 n}{4-3 n}+\frac{18}{4-3 n}-6\, .
\end{align}
By solving the differential equation (\ref{firstperturbativeorder}), we obtain the solution,
\begin{equation}\label{firstperturbationsolutioon}
F_1(\mathcal{G})=\frac{\Gamma  \mathcal{G}^{\delta }}{(\delta -1) (\delta  K-1)}\, ,
\end{equation}
therefore the perturbation expansion (\ref{fg3}) at linear order in the parameter $\varepsilon$ is,
\begin{equation}\label{fullfrfirstorder}
F(\mathcal{G})=-\frac{B \mathcal{G}^{-\gamma }}{(\gamma +1) (\gamma  K+1)}+c_1 \mathcal{G}^{1/K}+c_2 \mathcal{G}+\varepsilon \frac{\Gamma  \mathcal{G}^{\delta }}{(\delta -1) (\delta  K-1)}\, ,
\end{equation}
and as we already mentioned, the quantum correction will not affect the classical result with regards to the spectral index, therefore the LQC $F(\mathcal{G})$ theory for the intermediate inflation scenario is not viable.

Things are different however for the case of the singular bounce. In this case the classical theory is viable, as was demonstrated in Ref. \cite{Oikonomou:2015qha}, so the same will apply for the LQC corrected theory. Let us now use the formalism we presented in the previous section in order to find the first LQC correction to the classical result. The singular bounce is described by the evolution (\ref{bambabounce}) but with $n>2$, in which case the singularity at the origin is a Type IV singularity. In this case, the Gauss-Bonnet invariant for early times can be approximated as follows,
\begin{equation}\label{approximategaussbonnet1}
\mathcal{G}=24 A^3 n^3 (n-1) t^{3 n-4}\, ,
\end{equation}
which means that since $n>2$, as $t \to 0$, $\mathcal{G}\to 0$. Therefore, in this case, the leading order contributions in $\mathcal{G}$ should be chosen by simply keeping dominant terms as $\mathcal{G}\to 0$. By doing so, the differential equation (\ref{fg4}) reads in this case,
\begin{equation}\label{diffeqnnewleadingordering1}
\frac{B}{\mathcal{G}^{\gamma }}+\mathcal{G}^2 \Gamma F_0''(\mathcal{G})-\mathcal{G} F_0'(\mathcal{G})+F_0(\mathcal{G})=0\, ,
\end{equation}
where in this case, the parameters $\gamma$, $B$ and $\Gamma$ are equal to,
\begin{align}\label{parametersBandK1}
& \gamma=\frac{2 (n-1)}{3 n-4},\\ \notag &
B=A^2 2^{\frac{7-3 n}{4-3 n}} 3^{\frac{1}{4-3 n}+1} \left(\frac{1}{A^3 (n-1)}\right)^{\frac{2 (n-1)}{3 n-4}},\\ \notag &
\Gamma=\frac {3 n - 4} {n - 1}\, .
\end{align}
The differential equation (\ref{diffeqnnewleadingordering1}) can be solved analytically and the solution is,
\begin{equation}\label{classicalcontribution1}
F_0(\mathcal{G})=-\frac{B \mathcal{G}^{\gamma }}{(\gamma -1) (\gamma  \Gamma -1)}+c_1 \mathcal{G}^{1/K}+c_2 \mathcal{G}\, ,
\end{equation}
which is the classical contribution to the LQC $F(\mathcal{G})$ gravity, where $K$ is given in Eq. (\ref{parametersBandK}). Accordingly, by solving the differential equation (\ref{firstperturbativeorder}), we obtain the following solution,
\begin{equation}\label{firstperturbationsolutioon1}
F_1(\mathcal{G})=\frac{\Gamma  \mathcal{G}^{\delta }}{(\delta -1) (\delta  K-1)}\, ,
\end{equation}
where $\delta$ stands for,
\begin{equation}\label{deltanewparmeter}
\delta=-3 \gamma -\frac{4 (n-1)}{4-3 n}-\frac{12 n}{4-3 n}+\frac{18}{4-3 n}-6\, ,
\end{equation}
therefore the perturbation expansion of Eq. (\ref{fg3}) at linear order in the perturbation parameter $\varepsilon$ is in this case,
\begin{equation}\label{fullfrfirstorder1}
F(\mathcal{G})=-\frac{B \mathcal{G}^{\gamma }}{(\gamma -1) (\gamma  \Gamma -1)}+c_1 \mathcal{G}^{1/K}+c_2 \mathcal{G}+\varepsilon \frac{\Gamma  \mathcal{G}^{\delta }}{(\delta -1) (\delta  K-1)}\, .
\end{equation}
Therefore, the LQC-corrected Gauss-Bonnet gravity which realizes the Type IV singular bounce is given in Eq. (\ref{fullfrfirstorder1}), and as it is anticipated the power spectrum is mainly affected by the classical contribution $F_0(\mathcal{G})$, since the first quantum correction has a subleading contribution to $n_s$, depending strongly on the perturbation parameter $\varepsilon$.

\section*{Conclusions}

In this work we investigated which LQC-corrected $F(\mathcal{G})$ gravity may realize the intermediate inflation and the Type IV singular bounce cosmologies. We use the formalism of holonomy corrected LQC $F(\mathcal{G})$ gravity, and for both the cosmological evolutions we found that the classical description agrees with the results coming from the related literature. The LQC-corrected $F(\mathcal{G})$ gravity has a perturbative expansion, in which the perturbation parameter is equal to the term $\frac{1}{\rho_c}$, where $\rho_c$ is the critical density. The perturbation expansion appears in powers of $\frac{1}{\rho_c}$, as the critical density $\rho_c$ tends to infinity. In this case we may have a direct idea of the first LQC corrections near the classical limit of the theory, which is achieved when $\rho_c\to \infty$. As it is expected, the first quantum corrections do not alter significantly the spectral index of the primordial curvature perturbations, in which case the classical picture suffices and is dominant. We expect this result to be true unless the parameter $\rho_c$ takes small values, in which case though the perturbative expansion would break down. Although we examined only two cosmological evolutions, we expect the result to be quite general, and it hold true for other viable cosmologies. We believe that there is only one case in which the quantum corrections would play some role, and this is for classical cosmologies which are nearly scale invariant. It might then be possible that the quantum corrections will provide a subleading and wanted blue tilt in the power spectrum.

\section*{Acknowledgments}

Financial support by the Research
Committee of the Technological Education Institute of Central Macedonia, Serres,
under grant SAT/ME/011117-193/13, is gratefully acknowledged.

\end{document}